# The origin of the unusual anisotropic electrical conductivity of $Ba_3Al_2As_4$

Gui Yang(1)(2), Guangbiao Zhang(1), Jueming Yang(1), Chao Wang(1), Yuanxu Wang (1) [*]

(1) *Institute for Computational Materials Science, School of Physics and Electronics, Henan University, Kaifeng 475004, China*

(2) *College of Physics & Electrical Engineering, Anyang Normal University, Anyang 455000 China*

Abstract: $Ba_3Al_2As_4$ exhibits an unusual anisotropic electrical conductivity, that is, the electrical conductivity along the chain is smaller than those along other two directions which conflicts previous conclusion. Earlier studies on $Ca_5M_2Pn_6$ showed a high electrical conductivity along the chain. The band decomposed charge density is used to explain such unusual behavior. The results indicate that the existence of a conductive pathway near the Fermi level is responsible for the electrons transport. Further, the Ba-As bonding of $Ba_3Al_2As_4$ is some degree covalency which is novel for the Zintl compounds.

# 1. Introduction

Thermoelectric materials have attracted much attention in scientific community due to their potential technological applications [1-8]. The thermoelectric performance is quantified by the figure of merit, $ZT = \dfrac{S^2 \sigma T}{\kappa_e + \kappa_l}$, where $S$ is the Seebeck coefficient, $\sigma$ is the electrical conductivity, $T$ is the absolute temperature, and $\kappa_e$, $\kappa_l$ is the electronic and lattice thermal conductivity, respectively.

Zintl compounds[9-11] are promising thermoelectric candidates due to its "phonon-glass and electron-crystal" characteristics, which are made up of two substructures (typically, Groups 1 and 2). The electropositive cation (Group 1) donates its electrons to the anionic species (Group 2), which, in turn, forms the covalent bonding. Generally, the covalent bonding in the anionic substructure allows a higher mobility of the charge-carrier species. One example was $Ca_5M_2Sb_6$ (M=Al, Ga, In) which showed, proven by the experiment[12], that a stronger covalent bonding between Ga-Sb benefitted to form a lighter band mass and higher mobility. Therefore, the electrical conductivity $\sigma$ of $Ca_5Ga_2Sb_6$ was the largest in $Ca_5M_2Sb_6$ compounds. However, the experiment has not studied the anisotropy transport properties. Meanwhile, theoretical calculations [13] on $Ca_5Al_2Sb_6$ also predicted that the electrical conductivity along the z direction (in the covalent bonding direction) was much higher than that along the x and y directions. However, a clear explanation about the high $\sigma$ along the covalent-bonding direction has not been obtained as far as we know.

Here, we focused on a Zintl compound $Ba_3Al_2As_4$ synthesized recently with the space group *Pnma* [14]. The electronic structure and the thermoelectric properties of $Ba_3Al_2As_4$ are studied by first-principles method and semiclassical Boltzmann theory[15]. $Ca_5Ga_2As_6$ [16] is a typical representative of $Ca_5M_2Pn_6$ which exhibits high thermoelectric performance. The comparison between $Ba_3Al_2As_4$ and $Ca_5Ga_2As_6$ shows that $Ba_3Al_2As_4$ exhibits an unusual anisotropic electrical conductivity which can be explained by the calculated band decomposed charge density.

The full potential linearized augmented plane-wave (LAPW) method as implemented in the WIEN2K[17] is used to relax the lattice structure and calculate the electronic structure of $Ba_3Al_2As_4$. The exchange-correlation function is the generalized-gradient approximation of Perdew, Burke, and Ernzerhof (PBE-GGA) [18]. Self-consistent calculations are carried out with $10\times10\times10$ mesh k points in the Brillouin zone, and much denser grids 18000 k points for the transport calculations. The transport properties based on the Fourier expansion of the band energies are estimated using the semiclassical Boltzmann theory[15]. The scattering time $\tau$ approximation is considered as a constant in the BoltzTrap code.

The crystal structure of $Ba_3Al_2As_4$ is shown in Fig. 1(a). Like the $Ca_5M_2Pn_6$ (M=Al, Ga, In, Pn=Sb, As)[12, 13, 16, 19-23], the lattice structure of $Ba_3Al_2As_4$ exhibits an anisotropic quasi-one-dimensional chain linked by the covalent Al-As bond. There are two crystallographically unique Ba atoms and three crystallographically unique As atoms labeled as Ba1, Ba2, As1, As2 and As3, respectively. The edge-shared tetrahedra form a quasi-one-dimensional zigzag chain structure along the y-direction. The optimized lattice parameters are a=7.553 Å, b=11.9578 Å, and c=11.9829 Å for $Ba_3Al_2As_4$, respectively. These relaxed lattice constants are close to the experimental values[14]. The detailed calculated results of $Ca_5Ga_2As_6$ can be referred to Ref.[16]. To have a clearer comparison, we recalculate the properties of $Ca_5Ga_2As_6$ and the results are provided in the support information(SI).

The electronic localization functions (ELF) revealing the binding of electrons is useful in distinguishing between metallic, covalent, and ionic bonding. The Al atom in Fig. 1(b) donates its valence electrons to the As atoms, and the electrons are mainly localized around the As atoms. On the other hand, ELF near the middle of Al-As bonding attains local maximum values. Therefore, it exhibits the combination of the stronger covalent and weaker ionic interactions between the Al and As atoms. The ELF is evenly distributed around the Ba atom with the ELF>0.8, and no obvious electrons localized in the middle of Ba and As. Therefore, the interaction between Ba atoms and As atoms is weak. The calculated crystal orbital Hamilton populations(COHP) in Ref.[14] showed that the weak interaction between the Ba and As atoms was some degree covalency. For the Ga-As contacts (SI-Fig. 1(b)), the electrons are distributed around the Ga and As atoms with the maximum ELF in the center of Ga-As. Thus, a strong covalent bonding between the Ga and As atoms is formed. We note that electrons also evenly distribute around the Ca atoms, and the ELF value is much smaller than the one around the Ba atoms. Thus, the charges transfer from the Ca atoms to the As atoms means an ionic Ga-As bonding. The different effects of Ba and Ca on chemical bonding suggest that the Zintl formalism over-simplifies the interactions between cations and anions, and the actual bonding pictures in these compounds could be far more complicated.

Fig. 2(a) shows the project density of states (PDOS) which agrees well with the results in Ref.[14]. We only focus on the PDOS near the Fermi level because it is strong related to the transport properties. It is clear in Fig. 2(a) that the upper valence bands of $Ba_3Al_2As_4$ mainly arise from the As atoms, and the conduction band bottom is mainly from the Ba atoms. For $Ca_5Ga_2As_6$ (SI-Fig. 2), it tells us that the edges of both the valence and the conduction bands are mainly derived from As orbitals. The band structures of $Ba_3Al_2As_4$ are shown in Fig. 2(b). The valence band maximum (VBM) and the conduction band minimum (CBM) are both located at Γ point. The direct band gap is 1.12 eV which is smaller than 1.3 eV with the tight-binding linear-muffin-tin-orbital (TB-LMTO-ASA) method[14].

The anisotropy of the thermoelectric transport of $Ba_3Al_2As_4$, as a function of carrier concentrations, including $S$, $\sigma/\tau$, and $S^2\sigma/\tau$ along the x, y, and z directions are plotted in Fig. 3. The rigid band approximation is used to calculate transport properties for various doping level. The positive and negative values represent *p*-type and *n*-type doping. Due to the similar results at different temperatures, we only show the anisotropy transport for $T$= 500 K in the left map. As known[12], the Seebeck coefficient is proportional to the density-of-states effective mass $m_{DOS}^* = m_b^* N_v^{2/3}$ and the band mass $m_b^* = \hbar^2 \left[ \frac{\partial^2 E}{\partial k^2} \right]_{E(k)=E_f}^{-1}$, while the electrical conductivity is inversely proportional to the band effective mass by the $\sigma = ne\mu$ and $\mu \propto \frac{1}{(m_b^*)^{5/2}}$. $N_v$ is the band degeneracy, $e$ is the electronic charge, $n$ is the carrier concentration and $\mu$ is the carrier mobility.

For a qualitative discussion, $m_{DOS}^*$ and $m_b^*$ are calculated based on the band structure (Fig. 2(b)). In the valence bands, the effective masses along different directions are $m_{\Gamma-X}^* = -8.48 m_e$, $m_{\Gamma-Y}^* = -16.47 m_e$, and $m_{\Gamma-Z}^* = -9.73 m_e$ with $m_e$ is the electron mass. The larger dispersion contributes to a larger $\sigma/\tau$, and the high degeneracy is in favor of a bigger $S$. Due to the high degeneracy along Γ-X direction, $m_{DOS}^* = -21.37 m_e$ makes a dominant contribution to the $S_x$. Therefore, $|m_{DOS}^*| > |m_{\Gamma-Y}^*| > |m_{\Gamma-Z}^*|$ leads to the result of $S_x > S_y > S_z$, and $|m_{\Gamma-Y}^*| > |m_{\Gamma-Z}^*| > |m_{\Gamma-X}^*|$ causes $(\sigma/\tau)_x > (\sigma/\tau)_z > (\sigma/\tau)_y$ in the *p*-type doping region as shown in Fig. 3(a) and Fig. 3(b). For the *n*-type doping, the effective mass is $m_{\Gamma-X}^* = 7.62 m_e$, $m_{\Gamma-Y}^* = 11.28 m_e$, and $m_{\Gamma-Z}^* = 11.4 m_e$, meaning the $S_y > S_z > S_x$ and $(\sigma/\tau)_x > (\sigma/\tau)_z > (\sigma/\tau)_y$. A further explanation about the electrical conductivity is discussed below.

It is obviously in Fig. 3(b) that the anisotropy of $\sigma/\tau$ with *n*-type doping is larger than that with *p*-type doing, and consequently the higher $S^2\sigma/\tau$ (Fig. 3(c)). The more important thing is that the largest $S$ and $\sigma/\tau$ along the x direction exists simultaneously in the *p*-type doping region. This is helpful to enhance the thermoelectric performance. Then, the temperature dependence of the transport properties along the x direction is considered in the right map. It can be seen in Fig. 3(d) and (f) that the $S$ and $S^2\sigma/\tau$ increase with increasing temperature. However, a bipolar effect occurs in Fig. 3(d) at *T*=1000 K which depresses the *S*. For Fig. 3(e), the decreasing $\sigma/\tau$ is mainly caused by the increased phonon scattering with increasing temperature. The anisotropy of the thermoelectric transport of $Ca_5Ga_2As_6$ is shown in SI-Fig.3. Obviously, the results are almost similar with that of $Ba_3Al_2As_4$. As known, bipolar effect is a consequence of a small band gap that generates two kinds of carriers participating in transport. Hence, a small band gap, 0.37eV of $Ca_5Ga_2As_6$[16] induces the stronger bipolar effect happened at lower temperature (500 K) than that for $Ba_3Al_2As_4$ (1000 K). Also, the comparison between two compounds shows that $Ba_3Al_2As_4$ has a high thermoelectric performance, and the n-type doping may achieve a higher thermoelectric performance than the p-type doping.

Finally, we note the electrical conductivity $\sigma/\tau$ along the quasi-one-dimensional chain in the lattice structure. For $Ca_5Ga_2As_6$ (SI-Fig.3), the $(\sigma/\tau)_z$ (in the chain direction) is much better than those along the x- and y-directions which is in agreement with previous studies on the $Ca_5M_2Pn_6$ compounds[12, 13]. For $Ba_3Al_2As_4$, the quasi-one-dimensional chain, formed by the connecting Al-As bond, is along the y-direction. But, the electrical conductivity $(\sigma/\tau)_y$ (Fig. 3(b)) does not exhibit a better performance; on the contrary it is smaller than those along the x- and z-directions. This is contradicted to behaviors for the $Ca_5M_2Pn_6$ system. As known, the transport properties are strongly related to the bands near the Fermi level. So, we calculate the band decomposed charge density to explain such interesting transport behaviors. The band decomposed charge density shows the real-space distribution of the corresponding electronic states, which has been proved to be an effective way to study the electrons transport [24].

Fig. 4 shows the charge density distributions of $Ba_3Al_2As_4$ in the given energy windows 0.0 to -0.3 eV near the VBM for (a) and (b), 1.12 to 1.42 eV near the CBM for (c) and (d). The electronic states near the VBM are mainly from the As2, As3 and As1 atoms, and the electronic states near the CBM are mainly from the Ba1 and Ba2 atoms. This is consistent with the results of Fig. 2(a). The isosurface value of Fig. 4(a) ranges from Fmin=0 to Fmax=0.007, and the displayed isosurface value is 0.001. The distribution of electronic states usually determines the electrons transport. We find that the charge densities mainly accumulate in two layers parallel to the x-z plane. Two layers correspond to Ba1-As2-Ba1 plane and Ba2-As3-Ba2 plane, respectively. These two layers are separated along the y-direction. Therefore, it is easy to form a conductive pathway in the layers which helps the electrons transport. The contour plot of the distribution of electronic states in the Ba1-As2-Ba1 plane is plotted in Fig. 4(b). Apparently, the distribution of electronic states along the x-direction is more concentrated than that along the z-direction. Thus, the conductive pathway along the x-direction is more conducive to the electrons transport than that along the z-direction. That is why the $(\sigma/\tau)_x > (\sigma/\tau)_z > (\sigma/\tau)_y$ for the p-type doping $Ba_3Al_2As_4$. In Fig. 4(c) and (d), the isofurface value is in the range of 0 to 0.0043, and the displayed isosurface value is fixed at 0.001. Along the x-direction, there is a distinct conductive pathway formed by the connected Ba1-Ba1 atoms, which is responsible for the electrons transport. Thus, the $(\sigma/\tau)_x$ is much higher than $(\sigma/\tau)_y$ and $(\sigma/\tau)_z$. The same reason is used to explain the higher $(\sigma/\tau)_z$ in the $Ca_5Ga_2As_6$ and the detailed description is listed in the SI. The calculated charge densities in different energy windows show that the Ba and Ca cations have a great contribution to the electrons transport. The current study will provide a more insight into the transport properties in the Zintl compounds.

In summary, electronic structure and thermoelectric properties of $Ba_3Al_2As_4$ are studied by using ab initio calculations and semiclassical Boltzmann theory. The ELF shows that it has some degree of covalency of the Ba-As bonding, which is novel for the Zintl compounds. Further, the calculated anisotropic transport properties show that the $\sigma/\tau$ of $Ba_3Al_2As_4$ exhibits unusual transport behaviors along the chain, which contradicts previous studies on the $Ca_5M_2Pn_6$ compounds[12, 13]. The reason is mainly attributed to the distribution of the electronic states near the Fermi level. The formation of the conductive pathways results in a higher electrical conductivity.

**Acknowledgments**

**References**

[1] Y. Wang, X. Chen, T. Cui, Y. Niu, Y. Wang, M. Wang, Y. Ma, and G. Zou, Phys. Rev. B **76**, 155127 (2007).

[2] J. R. Sootsman, D. Y. Chung, and M. G. Kanatzidis, Angew. Chem. Int. Ed. **48**, 8616 (2009).

[3] J. R. Sootsman, H. Kong, C. Uher, J. J. D'Angelo, C.-I. Wu, T. P. Hogan, T. Caillat, and M. G. Kanatzidis, Angew. Chem. Int. Ed. **47**, 8618 (2008).

[4] K. Yang, Y. Chen, R. D'Agosta, Y. Xie, J. Zhong, and A. Rubio, Phys. Rev. B **86**, 045425 (2012).

[5] H. W. Leite Alves, A. R. R. Neto, L. M. R. Scolfaro, T. H. Myers, and P. D. Borges, Phys. Rev. B **87**, 115204 (2013).

[6] V. Pardo, A. S. Botana, and D. Baldomir, Phys. Rev. B **87**, 125148 (2013).

[7] D. Parker, X. Chen, and D. J. Singh, Phys. Rev. Lett. **110**, 146601 (2013).

[8] H. S. Ji, H. Kim, C. Lee, J.-S. Rhyee, M. H. Kim, M. Kaviany, and J. H. Shim, Phys. Rev. B **87**, 125111 (2013).

[9] Slack, G. A. CRC Handbook of Thermoelectrics, Rowe, D. M.; CRC Press: Boca Raton, FL, (1995).

[10] G. S. Nolas, J. Poon, M. Kanatzidis, Mater. Res. Soc. Bull. **31**, 199-205(2006).

[11] S. M. Kauzlarich, S. R. Brown, G. J. Snyder, Dalton Trans. 2099 (2007).

[12] A. Zevalkink, G. S. Pomrehn, S. Johnson, J. Swallow, Z. M. Gibbs, and G. J. Snyder, Chem. Mater. **24**, 2091 (2012).

[13] Y. L. Yan and Y. X. Wang, J. Mater. Chem. **21**, 12497 (2011).

[14] H. He, C. Tyson, M. Saito, and S. Bobev, J. Solid State Chem. **188**, 59 (2012).

[15] G. K. H. Madsen and D. J. Singh, Comput. Phys. Commun. **175**, 67 (2006).

[16] Y. L. Yan, Y. X. Wang, and G. B. Zhang, J. Mater. Chem. **22**, 20284 (2012).

[17] P. Blaha, K. Schwarz, G. Madsen, D. Kvasnicka, J. Luitz, WIEN2k: an Augmented Plane Wave Plus Local Orbitals Program for Calculating Crystal Properties; Institute of Physical and Theoretical Chemistry: Vienna, Austria, WIEN2k (2001)

[18] J. P. Perdew, K. Burke, M. Ernzerhof, Phys. Rev. Lett. **77,** 3865 (1996).

[19] E. Flage-Larsen, S. Diplas, Ø. Prytz, E. S. Toberer, and A. F. May, Phys. Rev. B **81**, 205204 (2010).

[20] E. S. Toberer, A. Zevalkink, N. Crisosto, and G. J. Snyder, Adv. Funct. Mater. **20**, 4375 (2010).

[21] L. Zhang, M.-H. Du, and D. J. Singh, Phys. Rev. B **81**, 075117 (2010).

[22] A. Zevalkink, E. S. Toberer, W. G. Zeier, E. Flage-Larsen, and G. J. Snyder, Energy & Environmental Science **4**, 510 (2011).

[23] W. G. Zeier, A. Zevalkink, E. Schechtel, W. Tremel, and G. J. Snyder, J. Mater. Chem. **22**, 9826 (2012).

[24] L. Xi, Y. B. Zhang, X. Y. Shi, J. Yang, X. Shi, L. D. Chen, W. Zhang, J. Yang, D. J. Singh, Phys. Rev. B, 86, 155201 (2012).

**Figure caption**

Figure 1. (Color online) (a) The lattice structure of $Ba_3Al_2As_4$. (b) The calculated ELF with the isosurface ELF=0.8.

Figure 2. (Color online) The calculated PDOS and band structure of $Ba_3Al_2As_4$.

Figure 3. The anisotropic thermoelectric transport of $Ba_3Al_2As_4$ as a function of carrier concentration (units in cm$^{-3}$). (a) Seebeck coefficients, $S$ (unit in μV/k) (b) Electrical conductivities relative to relaxation time, $\sigma/\tau$ (unit in $10^{19}$ 1/Ω ms) (c) Powerfactor with respect to relaxationtime, $S^2\sigma/\tau$ (unit in $10^{11}$ W/K$^2$ms).

Figure 4. (Color online) Band decomposed charge density of $Ba_3Al_2As_4$ with the unit of charge density is e/Å$^3$. (a) The distribution of electronic states in the energy window -0.3 to 0 eV. (b) Contour plot on the Ba1-As2-Ba1 plane. (c) The distribution of electronic states in the energy window 1.12 to 1.42 eV. (d) Contour plot on the Ba1-As2-Ba1 plane.

Fig. 1

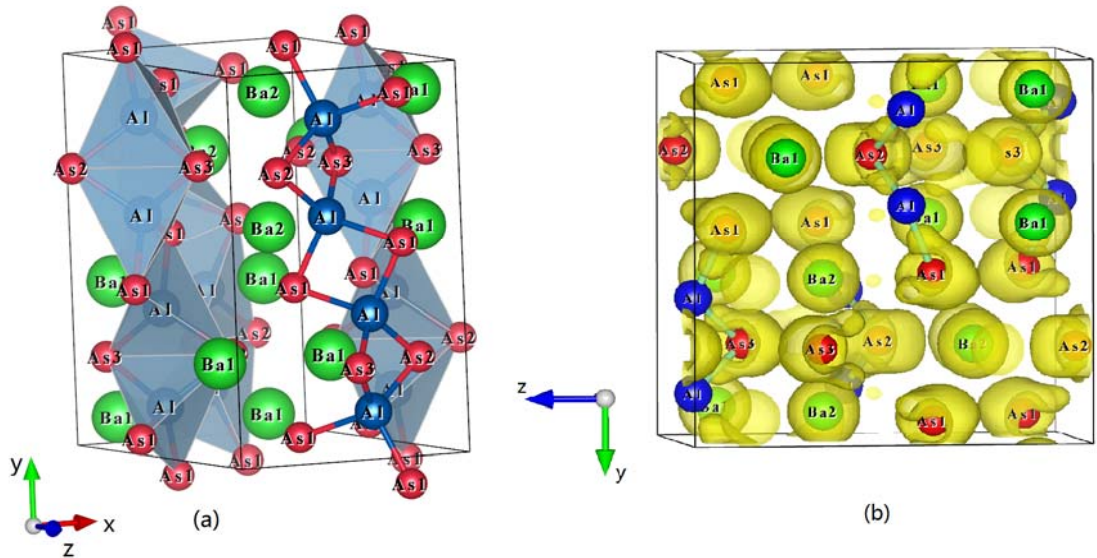

Fig. 2

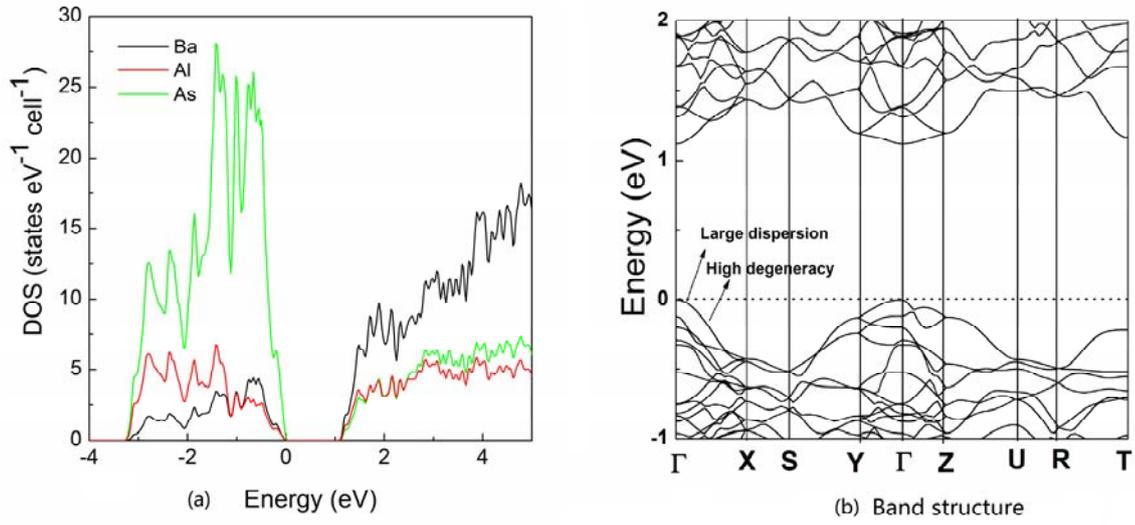

Fig. 3

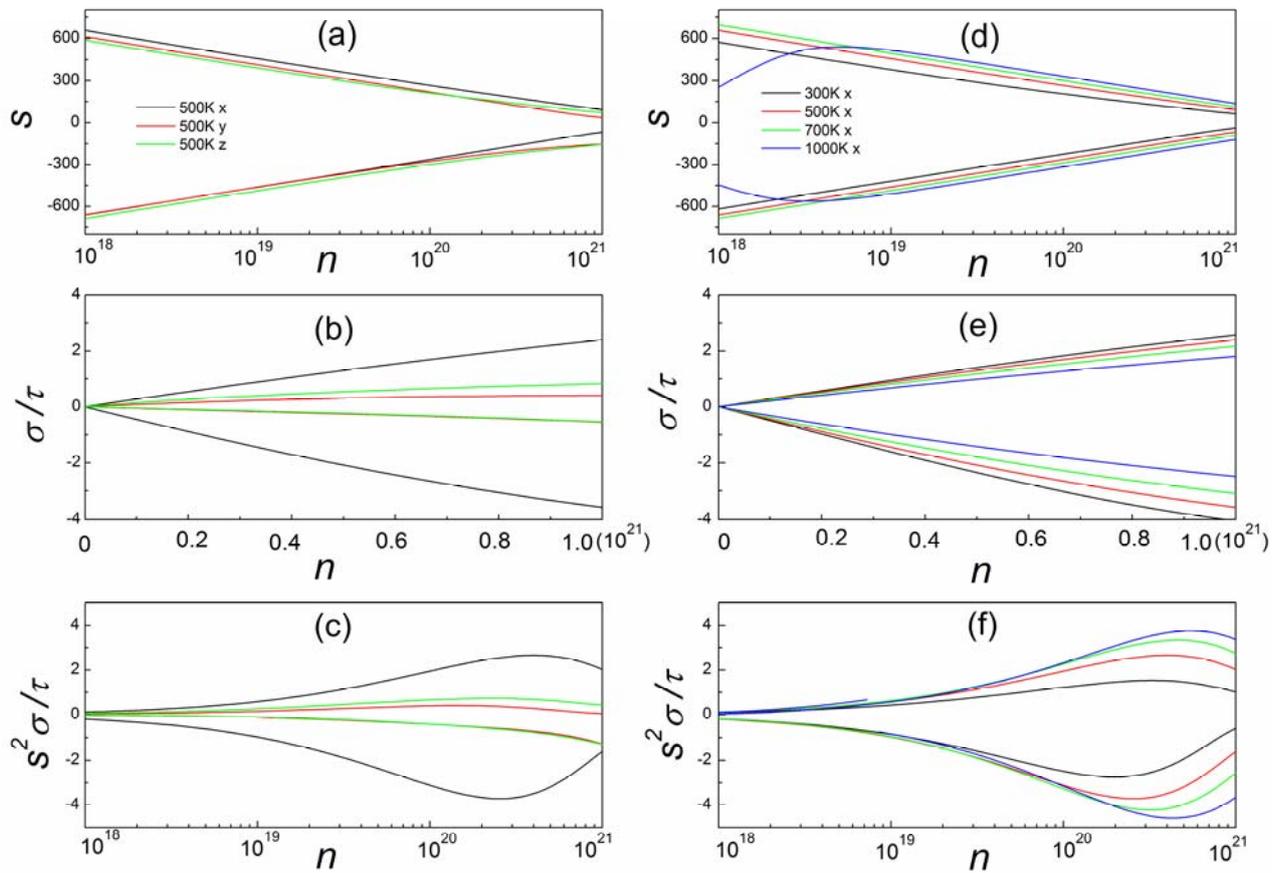

Fig.4

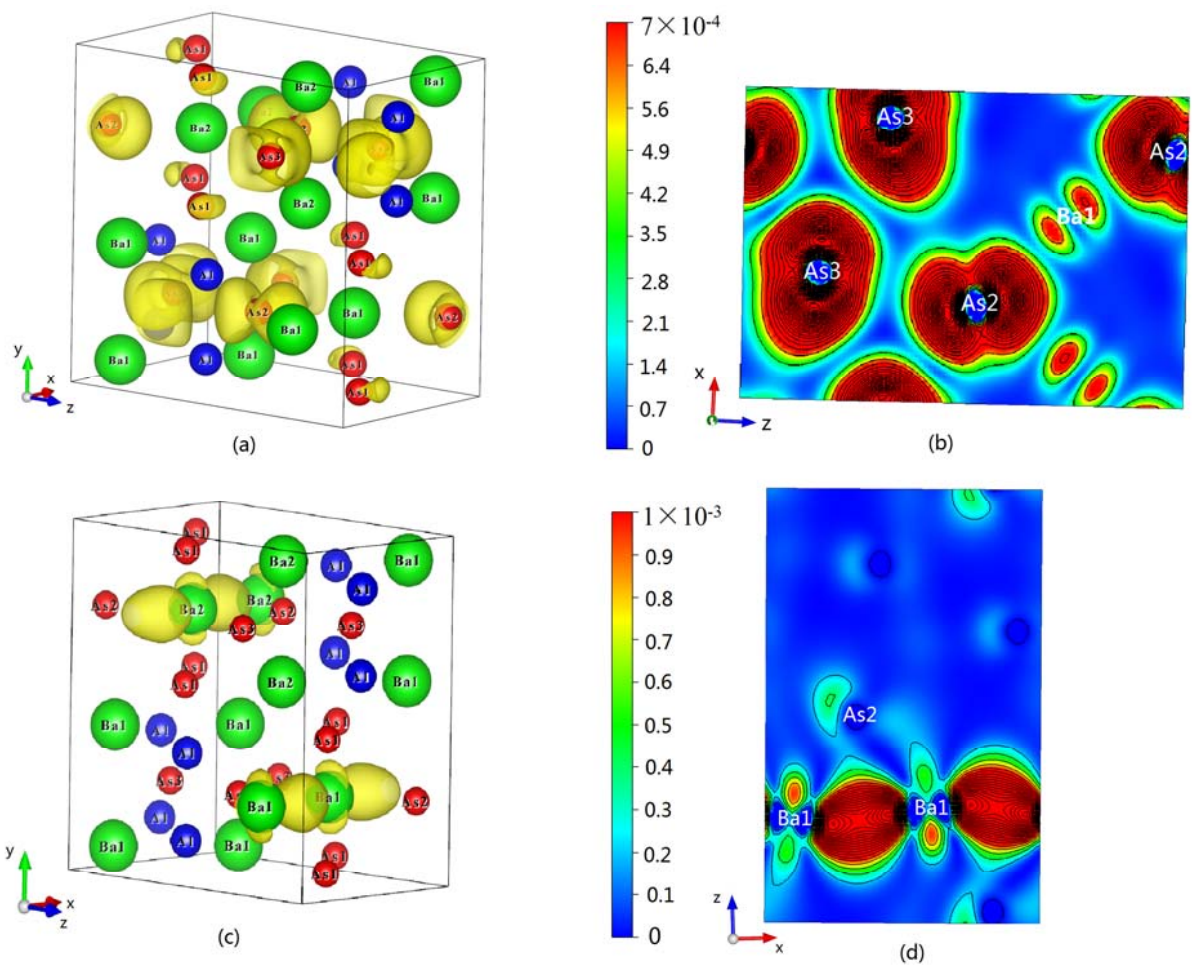